\input harvmac
\noblackbox
\newcount\figno
\figno=0
\def\fig#1#2#3{
\par\begingroup\parindent=0pt\leftskip=1cm\rightskip=1cm\parindent=0pt
\baselineskip=11pt
\global\advance\figno by 1
\midinsert
\epsfxsize=#3
\centerline{\epsfbox{#2}}
\vskip 12pt
\centerline{{\bf Figure \the\figno:} #1}\par
\endinsert\endgroup\par}
\def\figlabel#1{\xdef#1{\the\figno}}

\def\np#1#2#3{Nucl. Phys. {\bf B#1} (#2) #3}
\def\pl#1#2#3{Phys. Lett. {\bf B#1} (#2) #3}
\def\prl#1#2#3{Phys. Rev. Lett. {\bf #1} (#2) #3}
\def\prd#1#2#3{Phys. Rev. {\bf D#1} (#2) #3}


\font\cmss=cmss10
\font\cmsss=cmss10 at 7pt
\def\rlx{\relax\leavevmode}
\def\inbar{\vrule height1.5ex width.4pt depth0pt}
\def\IC{\relax\,\hbox{$\inbar\kern-.3em{\rm C}$}}
\def\IN{\relax{\rm I\kern-.18em N}}
\def\IP{\relax{\rm I\kern-.18em P}}
\def\ZZ{\rlx\leavevmode\ifmmode\mathchoice{\hbox{\cmss Z\kern-.4em Z}}
 {\hbox{\cmss Z\kern-.4em Z}}{\lower.9pt\hbox{\cmsss Z\kern-.36em Z}}
 {\lower1.2pt\hbox{\cmsss Z\kern-.36em Z}}\else{\cmss Z\kern-.4em
 Z}\fi}
\def\IZ{\relax\ifmmode\mathchoice
{\hbox{\cmss Z\kern-.4em Z}}{\hbox{\cmss Z\kern-.4em Z}}
{\lower.9pt\hbox{\cmsss Z\kern-.4em Z}}
{\lower1.2pt\hbox{\cmsss Z\kern-.4em Z}}\else{\cmss Z\kern-.4em
Z}\fi}

\def\narrowplus{\kern -.04truein + \kern -.03truein}
\def\narrowminus{- \kern -.04truein}
\def\narrowminussub{\kern -.02truein - \kern -.01truein}

\def\a{{\alpha}}

\def\r{{\rightarrow}}

\def\frac#1#2{{#1\over #2}}

\def\IZ{\relax\ifmmode\mathchoice
{\hbox{\cmss Z\kern-.4em Z}}{\hbox{\cmss Z\kern-.4em Z}}
{\lower.9pt\hbox{\cmsss Z\kern-.4em Z}}
{\lower1.2pt\hbox{\cmsss Z\kern-.4em Z}}\else{\cmss Z\kern-.4em
Z}\fi}
\def\IB{\relax{\rm I\kern-.18em B}}
\def\IC{{\relax\hbox{$\inbar\kern-.3em{\rm  C}$}}}
\def\ID{\relax{\rm I\kern-.18em D}}
\def\IE{\relax{\rm I\kern-.18em E}}
\def\IF{\relax{\rm I\kern-.18em F}}
\def\IG{\relax\hbox{$\inbar\kern-.3em{\rm G}$}}
\def\IGa{\relax\hbox{${\rm I}\kern-.18em\Gamma$}}
\def\IH{\relax{\rm I\kern-.18em H}}
\def\II{\relax{\rm I\kern-.18em I}}
\def\IK{\relax{\rm I\kern-.18em K}}
\def\IP{\relax{\rm I\kern-.18em P}}

\font\cmss=cmss10 \font\cmsss=cmss10 at 7pt
\def\IR{\relax{\rm I\kern-.18em R}}

\def\s{{\sigma}}
\def\1{{\bf 1}}
\def\3{{\bf 3}}
\def\7{{\bf 7}}
\def\2{{\bf 2}}
\def\8{{\bf 8}}

%

%
%
\def\eqnn#1{\xdef #1{(\secsym\the\meqno)}\writedef{#1\leftbracket#1}%
\global\advance\meqno by1\wrlabeL#1}
\def\eqna#1{\xdef #1##1{\hbox{$(\secsym\the\meqno##1)$}}
\writedef{#1\numbersign1\leftbracket#1{\numbersign1}}%
\global\advance\meqno by1\wrlabeL{#1$\{\}$}}
\def\eqn#1#2{\xdef #1{(\secsym\the\meqno)}\writedef{#1\leftbracket#1}%
\global\advance\meqno by1$$#2\eqno#1\eqlabeL#1$$}



\lref\rSS{S. Sethi and L. Susskind, hep-th/9702101,
    \pl{400}{1997}{265}.}
\lref\rBS{T. Banks and N. Seiberg,
    hep-th/9702187, \np{497}{1997}{41}.}
\lref\rreview{N. Seiberg, hep-th/9705117, Nucl. Phys. Proc. Suppl. 
{\bf 67} (1998) 158.}
\lref\rnahm{W. Nahm, \np{135}{1978}{149}.}
\lref\rnonpert{C. Montonen and D. Olive, \pl{72}{1977}{117}\semi
A. Sen, hep-th/9402032, \pl{329}{1994}{217}; hep-th/9402002,  
Int. J. Mod. Phys. {\bf
A9} (1994) 3707.}
\lref\rPD{J. Polchinski, hep-th/9510017, Phys. Rev. Lett. 
{\bf 75} (1995) 47.}
\lref\rWDB{E. Witten, hep-th/9510135, Nucl. Phys. {\bf B460} (1996) 335.}
\lref\red{E. Witten, hep-th/9710065, J.H.E.P. {\bf 1} (1998) 1.}
\lref\rhori{K. Hori, hep-th/9805141.}
\lref\rgimon{E. G. Gimon, hep-th/9806226.}
\lref\rkut{S. Elitzur, A. Giveon, D. Kutasov and D. Tsabar, hep-th/9801020, 
\np{524}{1998}{251}.}
\lref\rmald{J. Maldacena, hep-th/9711200.}
\lref\rbaryon{E. Witten, hep-th/9805112, J.H.E.P. {\bf 7} (1998) 6.}
\lref\rsenor{A. Sen, hep-th/9603113, Mod. Phys. Lett. {\bf A11} (1996) 1339.}
\lref\rprobe{N. Seiberg, hep-th/9606017, \pl{384}{1996}{81}.}
\lref\rvw{C. Vafa and E. Witten, hep-th/9505053, \np{447}{1995}{261}.}
\lref\rduff{M. Duff, J. Liu and R. Minasian, hep-th/9506126, \np{452}{1995}{261}.}
\lref\rsvw{S. Sethi, C. Vafa and E. Witten, hep-th/9606122, \np{480}{1996}{213}. }
\lref\rmukhi{K. Dasgupta, D. P. Jatkar and S. Mukhi, hep-th/9707224,
\np{523}{1998}{465}.}
\lref\rofer{O. Aharony, Y. Oz and Z. Yin, hep-th/9803051, \pl{430}{1998}{87}. }
\lref\rahn{C. Ahn, H. Kim and H. Yang, hep-th/9808182.}
\lref\rquant{E. Witten, hep-th/9609122, J. Geom. Phys. {\bf 22} (1997) 1.}


\lref\rpp{J. Polchinski and P. Pouliot, hep-th/9704029, \prd{56}{1997}{6601}.}
\lref\rds{M. Dine and N. Seiberg, hep-th/9705057, \pl{409}{1997}{239}.}
\lref\rdorey{N. Dorey, V. Khoze and M. Mattis, hep-th/9704197, 
\np{502}{1997}{94}.}
\lref\rbf{T. Banks, W. Fischler, N. Seiberg and L. Susskind, hep-th/9705190, 
\pl{408}{1997}{111}.}

\lref\rK{N. Ishibashi, H. Kawai, Y. Kitazawa and A. Tsuchiya, hep-th/9612115.}
\lref\rCallias{C. Callias, Commun. Math. Phys. {\bf 62} (1978), 213.}
\lref\rPD{J. Polchinski, hep-th/9510017, \prl{\bf 75}{1995}{47}.}
\lref\rWDB{E. Witten,  hep-th/9510135, Nucl. Phys. {\bf B460} (1996) 335.}
\lref\rSSZ{S. Sethi, M. Stern, and E. Zaslow, Nucl. Phys. {\bf B457} (1995)
484.}
\lref\rGH{J. Gauntlett and J. Harvey, Nucl. Phys. {\bf B463} 287. }
\lref\rAS{A. Sen, Phys. Rev. {\bf D53} (1996) 2874; Phys. Rev. {\bf D54} (1996)
2964.}
\lref\rWI{E. Witten, Nucl. Phys. {\bf B202} (1982) 253.}
\lref\rPKT{P. K. Townsend, Phys. Lett. {\bf B350} (1995) 184.}
\lref\rWSD{E. Witten, Nucl. Phys. {\bf B443} (1995) 85.}
\lref\rASS{A. Strominger, Nucl. Phys. {\bf B451} (1995) 96.}
\lref\rBSV{M. Bershadsky, V. Sadov, and C. Vafa, Nucl. Phys. {\bf B463}
(1996) 420.}
\lref\rBSS{L. Brink, J. H. Schwarz and J. Scherk, Nucl. Phys. {\bf B121}
(1977) 77.}
\lref\rCH{M. Claudson and M. Halpern, Nucl. Phys. {\bf B250} (1985) 689.}
\lref\rSM{B. Simon, Ann. Phys. {\bf 146} (1983), 209.}
\lref\rGJ{J. Glimm and A. Jaffe, {\sl Quantum Physics, A Functional Integral
Point of View},
Springer-Verlag (New York), 1981.}
\lref\rADD{ U. H. Danielsson, G. Ferretti, B. Sundborg, Int. J. Mod. Phys. {\bf
A11} (1996) 5463\semi   D. Kabat and P. Pouliot, Phys. Rev. Lett. {\bf 77}
(1996), 1004.}
\lref\rDKPS{ M. R. Douglas, D. Kabat, P. Pouliot and S. Shenker,
hep-th/9608024,
Nucl. Phys. {\bf B485} (1997), 85.}
\lref\rhmon{S. Sethi and M. Stern, Phys. Lett. {\bf B398} (1997), 47.}
\lref\rBFSS{T. Banks, W. Fischler, S. H. Shenker, and L. Susskind,
Phys. Rev. {\bf D55} (1997) 5112.}
\lref\rBHN{ B. de Wit, J. Hoppe and H. Nicolai, Nucl. Phys. {\bf B305}
(1988), 545\semi
B. de Wit, M. M. Luscher, and H. Nicolai, Nucl. Phys. {\bf B320} (1989),
135\semi
B. de Wit, V. Marquard, and H. Nicolai, Comm. Math. Phys. {\bf 128} (1990),
39.}
\lref\rT{ P. Townsend, Phys. Lett. {\bf B373} (1996) 68.}
\lref\rLS{L. Susskind, hep-th/9704080.}
\lref\rFH{J. Frohlich and J. Hoppe, hep-th/9701119.}
\lref\rAg{S. Agmon, {\it Lectures on Exponential Decay of Solutions of
Second-Order Elliptic Equations}, Princeton University Press (Princeton) 1982.}
\lref\rY{P. Yi, hep-th/9704098.}
\lref\rDLhet{ D. Lowe, hep-th/9704041.}
\lref\rqm{M. Claudson and M. Halpern, \np{250}{1985}{689}\semi
R. Flume, Ann. Phys. {\bf 164} (1985) 189\semi
M. Baake, P. Reinecke and V. Rittenberg, J. Math. Phys. {\bf 26} (1985) 1070.}
\lref\rbb{K. Becker and M. Becker, hep-th/9705091, \np{506}{1997}{48}\semi
K. Becker, M. Becker, J. Polchinski and A. Tseytlin, hep-th/9706072,
\prd{56}{1997}{3174}.}
\lref\rss{S. Sethi and M. Stern, hep-th/9705046. }
\lref\rpw{J. Plefka and A. Waldron, hep-th/9710104, \np{512}{1998}{460}.}
\lref\rhs{M. Halpern and C. Schwartz, hep-th/9712133.}
\lref\rlimit{N. Seiberg hep-th/9710009, \prl{79}{1997}{3577}\semi
A. Sen, hep-th/9709220.}
\lref\rentin{D.-E. Diaconescu and R. Entin, hep-th/9706059,
\prd{56}{1997}{8045}.}
\lref\rgreen{M. B. Green and M. Gutperle, hep-th/9701093, \np{498}{1997}{195}.}
\lref\rpioline{B. Pioline, hep-th/9804023.}
\lref\rgl{O. Ganor and L. Motl, hep-th/9803108.}
\lref\rds{M. Dine and N. Seiberg, hep-th/9705057, \pl{409}{1997}{209}.}
\lref\rberg{E. Bergshoeff, M. Rakowski and E. Sezgin, \pl{185}{1987}{371}.}
\lref\rBHP{M. Barrio, R. Helling and G. Polhemus, hep-th/9801189.}
\lref\rper{P. Berglund and D. Minic, hep-th/9708063, \pl{415}{1997}{122}.}
\lref\rspin{P. Kraus, hep-th/9709199, \pl{419}{1998}{73}\semi
J. Harvey, hep-th/9706039\semi
J. Morales, C. Scrucca and M. Serone, hep-th/9709063, \pl{417}{1998}{233}.}
\lref\rdine{M. Dine, R. Echols and J. Gray, hep-th/9805007.}
\lref\rber{D. Berenstein and R. Corrado, hep-th/9702108, \pl{406}{1997}{37}.}

\lref\rpss{S. Paban, S. Sethi and M. Stern, hep-th/9805018.}
\lref\rpsst{S. Paban, S. Sethi and M. Stern, hep-th/9806028. }
\lref\rperiwal{V. Periwal and R. von Unge, hep-th/9801121.}
\lref\rfer{M. Fabbrichesi, G. Ferreti and R. Iengo, hep-th/9806018.}


\Title{\vbox{\hbox{hep-th/9809162}
\hbox{IASSNS--HEP--98/78}}}
{\vbox{\centerline{A Relation Between N=8 Gauge Theories}
\vskip8pt\centerline{in Three Dimensions}}}
\centerline{ Savdeep
Sethi\footnote{$^\ast$} {sethi@sns.ias.edu} }
\medskip\centerline{\it School of Natural Sciences}
\centerline{\it Institute for Advanced Study}\centerline{\it
Princeton, NJ 08540, USA}

\vskip 0.5in

We show that three-dimensional N=8 $Sp(N)$ and $SO(2N+1)$ gauge
theories flow to the same 
strong coupling fixed point. As a consequence, the corresponding orientifold 
two-planes in type IIA string theory are described at strong coupling and 
low-energies by the same M theory background. 
In the large $N$ limit, these assertions
are confirmed by studying discrete torsion in the 
supergravity theory corresponding to membranes on $\IR^8/\IZ_2$.

\vskip 0.1in
\Date{9/98}

\newsec{N=8 Gauge Theories in Three Dimensions}

In four dimensions, N=4 Yang-Mills theories provide some of the simplest examples
of interacting conformal field theories. The existence of S duality implies that
there are at least two distinct realizations of the same conformal field 
theory \rnonpert.
The purpose of this letter is to explain a similar identification between conformal
field theories in three dimensions. Our starting point is N=8 Yang-Mills theory
in three dimensions with gauge group $G$.\foot{Both four-dimensional N=4
Yang-Mills and three-dimensional N=8 Yang-Mills have sixteen real supersymmetries.} 
The coupling constant, $g^2$, has dimension one in three dimensions. For simplicity,
we will assume there is only one coupling constant. To obtain an 
interacting conformal field theory,
we must therefore consider the infra-red limit of the theory where $g^2 \r\infty$
so the theory is strongly coupled.

Let the gauge group $G$ have rank $N$. The gauge theory contains seven scalar
fields, $\phi^i$,  in the adjoint representation of $G$. The moduli space of the
theory is, 
\eqn\moduli{ {\cal M} = { \IR^{7N} \times \widehat{T}^N \over {\cal W}},}
where $\cal W$ is the Weyl group of $G$ and $ \widehat{T}^N$ is the Cartan torus
for the dual group $\widehat G$ \rreview. The manifest R-symmetry of the theory is
$Spin(7)$. The compact directions in the moduli space 
$\cal M$ correspond to the expectation values of the scalar fields, $\s$, dual to 
the $N$ massless photons present at generic points in $\cal M$. The size of the compact
directions in $\cal M$ is proportional 
to $g^2$ and in the strong coupling limit, $\widehat{T}^N \r \IR^N $.

A non-trivial conformal field theory can only appear at a singularity of the 
moduli space $ \cal M$ in the strong coupling limit. For the case where
$G = U(N)$, arguments given in \refs{\rSS,
\rBS, \rreview} showed that the theory at the origin of the moduli space flows to 
an interacting $Spin(8)$ invariant fixed point. Note that there are other 
singularities in the moduli space but we will focus on the theory at the
origin. 

There are two ways to show $Spin(8)$ invariance. One argument only involves 
supersymmetry and the
conformal group in three dimensions, which is isomorphic to $Spin(3,2)$. Closure
of the superconformal algebra with sixteen supersymmetry generators simply 
requires a
$Spin(8)$ symmetry \refs{\rnahm, \rreview}. This argument is true for any
gauge group $G$. 

The second argument for $Spin(8)$ invariance will provide additional information. This
argument uses the strong-weak coupling duality of four-dimensional gauge theory.
Let us start with N=4 Yang-Mills in four dimensions with gauge group $G$ and 
coupling constant $\tau$. This theory has a dual realization in terms of a theory
with gauge group $ \widehat{G}$ and coupling $ -1/\tau$. We can set the theta angle 
to zero for our purposes. We can then restrict to the
$\IZ_2$ subgroup of the $SL(2, \IZ)$ strong-weak coupling duality which relates the 
dual coupling $ \tilde\lambda$ to the original 
four-dimensional coupling constant $\lambda$ in the following way:
\eqn\invert{ \tilde\lambda  =  {2\pi \over \lambda}.}

Let us compactify the 
gauge theory on $ \IR^3\times S^1$. The effective three-dimensional coupling constant
is given by,
\eqn\three{ g^2 = {\lambda^2 \over 2\pi R}, }
where $R$ is the radius of the circle. The component $A$ of the four-dimensional
gauge field along $S^1$ gives an extra adjoint-valued scalar,
\eqn\newscalar{\phi_e  = {1\over 2\pi R} \int_{S^1} A.}
This scalar field has a compact moduli space with a size proportional to $1/R$. In the
$R \r 0$ limit with $ g^2$ held fixed, this scalar field becomes $Spin(7)$ symmetric 
with the six scalars of the four-dimensional theory. 

In addition, we have another $N$ scalars, $\phi_m$, coming from dualizing
the three-dimensional photons. The moduli space for these scalars, which 
correspond
to the choice of 't Hooft lines or magnetic Wilson lines along $S^1$,  is also compact
with a size again proportional to $1/R$. In the $R \r 0$ limit with $\lambda$ 
held fixed,
these scalars become rotationally symmetric with the remaining non-compact  scalars.
Electric-magnetic duality then exchanges  $G$
with $  \widehat{G}$ and  exchanges the $\phi_e$ and $\phi_m$ directions.       

When $G$ and $ \widehat G$ are different,\foot{We are only interested in differences at
the level of the Lie algebra rather than the global structure of the group. For 
example when $G=U(N)$,
this $\IZ_2$ action is a symmetry if we choose the self-dual value for the coupling 
\rSS.} this 
duality action is not a symmetry of the theory for any finite value of the 
three-dimensional coupling constant. Rather, the $ \IZ_2$ duality identifies one 
theory with a different theory. In the infra-red limit where $g^2 \r \infty$, 
we can combine this duality with the known $Spin(8)$ invariance. The $Spin(8)$
invariance includes rotation of $\phi_e$ into $\phi_m$ but does not exchange
$G$ with $ \widehat G$.  We can therefore  
conclude that gauge theories in three dimensions with either gauge group $G$ or 
gauge group $ \widehat{G}$ flow to the same $Spin(8)$ invariant superconformal field
theory.

\newsec{Orientifolds and Supergravity}

The argument given in section one is primarily interesting for the non-simply-laced
gauge groups $Sp(N)$ and $SO(2N+1)$. For any finite coupling constant, these two
theories are quite different but they flow in the infra-red to the same conformal
field theory at the origin of the moduli space. These gauge theories appear naturally in 
type IIA string theory as the 
low-energy excitations of $N$ D2-branes coincident with various orientifold two-planes.
Our starting point is then type IIA string theory on the orientifold $ \IR^7/\IZ_2$.
The gauge theory coupling for D2-branes on top of the orientifold plane is related to
the  string coupling $g_s$ and string
scale $M_s$ by,
\eqn\scales{ g^2 = g_s M_s.}
The $ \IZ_2$ action must act on the circle on which we reduce from M theory to type
IIA if the moduli space seen by a D2-brane probing the orientifold is to agree with
\moduli\ for the gauge theory \refs{\rsenor, \rprobe}. When compactified on, 
\eqn\space{ \IR^7 \times S^1 \over \IZ_2,}
the Chern-Simons interaction
in eleven-dimensional supergravity, 
\eqn\chern{ -{1\over 6} \int{ C \wedge G \wedge G},}
is only invariant if $C$ is invariant under the $\IZ_2$ action. 
The M theory lift of the orientifold is 
therefore the $\IZ_2$ orbifold \space. The gauge theory coupling can be expressed 
in terms of the eleven-dimensional scale, $M_{pl}$, 
and the size $R_{11}$ of the $S^1$ in \space: 
$$ g^2 = M_{pl}^3 R_{11}^2. $$ 
As usual, to restrict to the field theory modes on the branes, we take the limit 
$M_{pl} \r \infty$ holding fixed $g^2$. The strong coupling limit for the gauge 
theory then corresponds to decompactifying 
the circle on which we reduce from M theory to type IIA. 

We will distinguish between three kinds of orientifold two-planes. The gauge 
symmetry of $N$ D2-branes on top of the orientifold plane $O2^-$ is $SO(2N)$. 
We will count branes and charges on the quotient space. This O2-plane carries $-1/8$
units of membrane charge.  We can also have a stuck $1/2$
membrane at the orientifold fixed point. We will call this case $ \widetilde{O2}^+$. 
In this case, the gauge symmetry is $SO(2N+1)$
and the charge is $3/8$. Lastly, we will also consider $O2^+$ which corresponds to the
choice of Chan-Paton factors giving gauge group $Sp(N)$. This O2-plane has charge
$1/8$.\foot{We will not distinguish cases where the Ramond gauge-field has a 
non-trivial holonomy around the circle in \space. This holonomy was important
in the case of $O4$-planes studied in \refs{\rhori, \rgimon}. In those cases, the
limit where $R_{11} \r \infty$, which we will primarily study, was not necessarily 
smooth. For example, $O4^0$
only exists with a finite size circle. In this respect, the case of $O2$-planes 
is nicer since the gauge theory has a smooth strong coupling limit so we can take 
the limit where $R_{11} \r \infty$. In this limit, 
any holonomy involving the circle goes away. }  

Let us start with the M theory lift of $O2^-$. There are two singularities in 
\space\ at $0$ and $\pi$ on the circle. At strong coupling, the geometry around 
either singularity becomes
$ \IR^8/\IZ_2$ and the D2-branes become membranes which we can place at the
fixed point. To compute the membrane charge of this configuration, we can
consider M theory on $T^8/\IZ_2$. There is a net membrane charge in this theory
coming from the interaction \refs{\rvw, \rduff}, 
\eqn\anom{ - \int{C \wedge X_8(R)}, }
in M theory. The total membrane charge is given by -$\chi/24$ and must be 
cancelled to avoid a tadpole anomaly \rsvw. Although the space $T^8/\IZ_2$ is
singular, standard string theory technology for orbifolds can be used to show
that $ \chi/24 = 16$. Each fixed point, and consequently the $\IR^8/\IZ_2$ 
fixed point, is then a source of $-1/16$ units of membrane charge 
\refs{\rmukhi, \rsenor}.        

To count the distinct M theory configurations corresponding to the remaining 
O2-planes, we turn to the supergravity solutions
dual to these conformal field theories \rmald. The supergravity solutions for
the various O3-planes were
analyzed in \rbaryon, while membranes on the orbifold space $\IR^8/\IZ_2$
were mentioned in \rofer. The $SO(2N)$ $(2,0)$ theory was considered from a 
supergravity perspective in \rahn. 
M theory compactified on $AdS_4 \times \IR P^7$ is 
dual to the world-volume theory of membranes on $\IR^8/\IZ_2$ in the limit 
where $M_{pl} \r \infty$.  
We wish to count the number of distinct fluxes for the four-form $G$ on $\IR P^7$. 
This counts the number of distinct strong coupling limits for $O2$-planes. Note
that since $w_4 =0$ for $ \IR^8/\IZ_2$, no  half-integral
$G$ flux is possible \rquant.  

The counting of fluxes is then rather simple. We need to compute $H^4( \IR P^7,
 \IZ)$ which, using the results of \rbaryon, is $\IZ_2$. Therefore, there 
is only one possible choice of discrete torsion. This is a supergravity
confirmation of the argument in section one. The low-energy theories on the 
three orientifold 
two-planes, $O2^-, \widetilde{O2}^+$ and $O2^+$ must flow to two distinct
strong coupling conformal field theories. We know that $O2^-$ flows to the
case without torsion so $\widetilde{O2}^+$ and $O2^+$ must flow to the case
with torsion. 

In the case where we turn on discrete torsion, we can ask: by how much does the
membrane charge change? The charge shift is generated by the term \chern\ and is
determined by computing,
\eqn\shift{ -{1\over 2} \int_{\IR P^7} {G\over 2\pi} \wedge {C\over 2 \pi},}
where $G$ is the torsion class. To compute this integral, we will view $ \IR P^7$
as the boundary of a smooth eight-dimensional space $\cal M$.\foot{I am especially
grateful to 
E. Witten for a detailed explanation about how to compute this integral.} 
We can then evaluate, 
$$   -{1\over 2} \int_{\cal M} {G\over 2\pi} \wedge {G\over 2 \pi},$$
rather than \shift. Let us begin by recalling the construction of the 
Hopf fibration of $S^7$ over
$ \, \IC P^3$. The sphere $S^7$ is the locus of points in $ \, \IC^4 $ obeying, 
$$ |z_1|^2+ |z_2|^2 + |z_3|^2 + |z_4|^2 = 1, $$
where $ \vec{z} = (z_1,z_2,z_3,z_4)$ coordinatize  $ \, \IC^4 $. To obtain $ \, \IC P^3$
from $S^7$,
we quotient by the $U(1)$ action, 
\eqn\quot{ \vec{z} \sim e^{i \a} \vec{z}. }  
The fibers over $ \, \IC P^3$ are then circles. We can describe this 
construction in a slightly different way. Let us start with the total space
of the bundle $O(-1)$ over $ \, \IC P^3$. Let $w$ be a coordinate for the fiber,
which is a copy of $ \, \IC$. We obtain a smooth eight manifold by 
taking a disk $ |w| <1$ in each fiber over $ \, \IC P^3$. The boundary of this
space is $S^7$. We can mimick this construction for the case of $ \IR P^7$ by
taking the bundle $O(-2)$ rather than $O(-1)$. This gives us the eight manifold
$\cal M$.     

Since $ \IR P^7$ is orientable, we can associate a homology class in 
$H_3( \IR P^7, \IZ)$ to $G$ using Poincar\'e duality. This class can be
 represented by an $\IR P^3$ subspace of $\IR P^7$ which is the boundary
of a four-cycle $ \cal W$ in $\cal M$.  Let $ \pi: {\cal M} \r \, \IC P^3$ denote 
the projection map. The Hopf bundle $ \IR P^3 \r \, \IC P^1$
is compatible with the projection map $\pi$ in the sense that $ \pi ({\cal W})$ is 
a $ \,\IC P^1$ in $\,\IC P^3$. 

The connection $C$ for the torsion
class $G$ obeys, 
\eqn\hol{   \int_{\cal \partial W} {C \over 2 \pi} = {1\over 2},}
and therefore,
\eqn\newhol{  \int_{\cal W} {G \over 2 \pi} = {1\over 2}.}
We need to identify a class $G$ in $\cal M$ obeying \newhol. The Poincar\'e
dual of the base $ \, \IC P^3$ of the bundle $\cal M$ is a two-form $X$ 
satisfying, 
$$ \int_{\cal W} X \wedge X = -2. $$ 
We can then take $ G/2\pi =  - X^2/4$. The charge shift is then 
given by,
\eqn\final{ \eqalign{ -{1\over 2 (2\pi)^2} \int_{\cal M} G\wedge G &
= -{1\over 16} \int_{\cal M} X^4, \cr &= {1\over 4}.}}
The final integral is evaluated by integrating over one $X$ which restricts the
remaining integral to the base $ \, \IC P^3$,  which is then standard.  

Turning on the discrete torsion therefore shifts the membrane charge from
$-1/16$ to $3/16$. There is an amusing interpretation for this $1/4$ membrane
charge. $ \widetilde{O2}^+$ differs from $O2^-$ by a stuck $1/2$ membrane. In
studying the strong coupling description of $ \widetilde{O2}^+$, we see that
the orientifold plane splits into two singularities from \space. Apparently the $1/2$
stuck membrane also splits into two fluxes, each carrying $1/4$ unit of membrane 
charge. The case of $O2^+$ is a little different. The singularity at the origin
of \space\ has charge $3/16$ but the other singularity at $\pi$ should have
charge $-1/16$ since the total membrane charge for $O2^+$ is $1/8$.

\bigbreak\bigskip\bigskip\centerline{{\bf Acknowledgements}}\nobreak
It is my pleasure to thank D. Morrison, N. Seiberg, A. Sen and especially 
E. Witten for helpful  
conversations. I would also like to thank the Aspen Center for Physics for
hospitality during the completion of this project. This work is supported 
by the William Keck Foundation and by 
NSF grant PHY--9513835.

\listrefs
\bye